\documentclass[sigconf]{acmart}

\usepackage{anyfontsize}

\usepackage{listings}  
\usepackage{enumitem}

\usepackage{booktabs}
\newcommand{\system}{VayuBuddy}

\newtoggle{CommentsMode}

\toggletrue{CommentsMode}

\newcommand{\nLLMs}{7}
\newcommand{\nPrompts}{45}

\AtBeginDocument{%
  }

\setcopyright{acmlicensed}
\copyrightyear{2024}
\acmYear{2024}

\acmConference[Conference acronym ’XX]{Make sure to enter the correct
  conference title from your rights confirmation emai}{June 03–05, 2018, Woodstock, NY}

\begin{document}


\newcommand{\ourtitle}{VayuBuddy: an LLM-Powered Chatbot to Democratize Air Quality Insights}
\title[\ourtitle]{\ourtitle}
\author{Zeel B Patel}
\email{patel_zeel@iitgn.ac.in}
\affiliation{%
  \institution{IIT Gandhinagar}
  \country{India}
}

\author{Yash Bachwana}
\email{yash.bachwana@iitgn.ac.in}
\affiliation{%
  \institution{IIT Gandhinagar}
  \country{India}
}


\author{Nitish Sharma}
\email{nitishsharma1295@gmail.com}
\affiliation{%
  \institution{Independent Researcher}
  \country{India}
}

\author{Sarath Guttikunda}
\email{sguttikunda@gmail.com}
\affiliation{%
\institution{UrbanEmmissions.info}
  \country{India}
}

\author{Nipun Batra}
\email{nipun.batra@iitgn.ac.in}
\affiliation{%
\institution{IIT Gandhinagar}
  \country{India}
}

\begin{abstract}
Nearly 6.7 million lives are lost due to air pollution every year. While policymakers are working on the mitigation strategies, public awareness can help reduce the exposure to air pollution. Air pollution data from government-installed-sensors is often publicly available in raw format, but there is a non-trivial barrier for various stakeholders in deriving meaningful insights from that data. In this work, we present \system, a Large Language Model (LLM)-powered chatbot system to reduce the barrier between the stakeholders and air quality sensor data. \system\ receives the questions in natural language, analyses the structured sensory data with a LLM-generated Python code and provides answers in natural language. We use the data from Indian government air quality sensors. We benchmark the capabilities of \nLLMs\ LLMs on \nPrompts\ diverse question-answer pairs prepared by us. Additionally, \system\ can also generate visual analysis such as line-plots, map plot, bar charts and many others from the sensory data as we demonstrate in this work.
\end{abstract}

\begin{teaserfigure}
  \includegraphics[width=\textwidth]{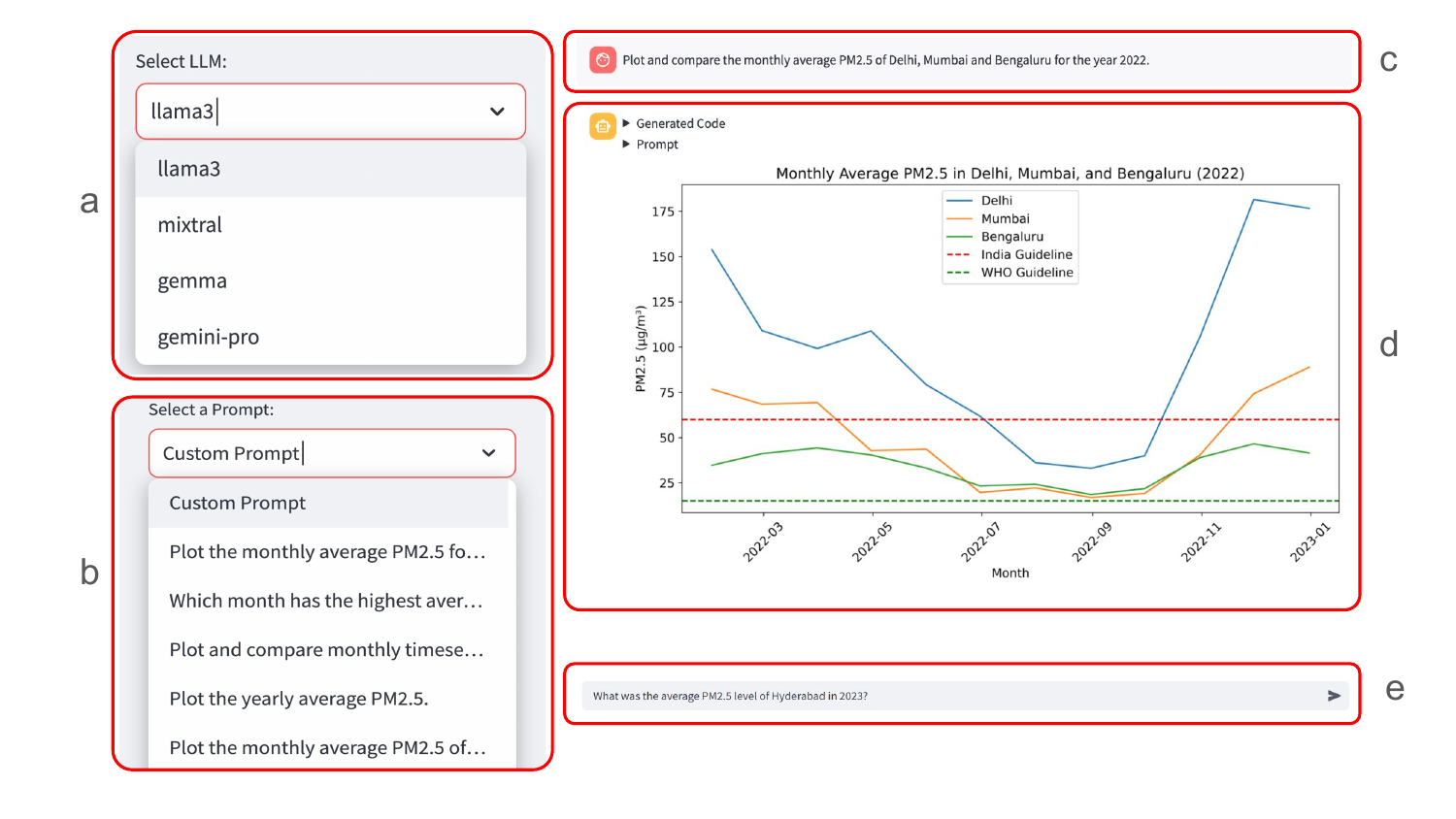}
  \caption{\system's Interface: a) selection of an LLM; b) selecting a prompt; c) question asked by a user; d) output from \system; e) chat input box to ask the questions to \system.
  }
  \Description{The Homepage of the \system web application.}
  \label{App Screenshot}
\end{teaserfigure}
\maketitle


Air pollution is one of the most critical environmental health risks of our time, silently killing over 6.7 million people annually~\cite{unep19}. 
Fine particulate matter pollution (PM$_{2.5}$) is a primary pollutant responsible for chronic obstructive pulmonary disease, lower respiratory infections, stroke, ischemic heart disease, and various cancers. Additionally, it plays a significant role in type 2 diabetes and neonatal disorders\footnote{\href{https://www.unep.org/interactives/air-pollution-note/}{https://www.unep.org/interactives/air-pollution-note/}}. Public/Media awareness about air pollution is limited since they are not equipped with the skills to process raw air pollution data to generate insights. 
Moreover, a growing literature~\cite{airpollcomm} on environmental health literacy suggests that communication about environmental risks must move beyond individual behavior education to empower communities to mobilize to reduce environmental threats. 
Media sources tend to present misperception and distortions regarding air quality risks, leading to a public disconnect from reality ~\cite{cisneros2017understanding}. Moreover, entertainment media exaggerates risks, contributing to misinformation~\cite{frayling2013mad}.

In recent years, advancements to transformer~\cite{vaswani2023attention} based large language models (LLMs) have revolutionized information retrieval and processing. Models like GPT~\cite{brown2020language}, BERT~\cite{devlin2019bert}, and Llama~\cite{touvron2023llama} leverage the transformer architecture to capture long-range dependencies, processing complex questions to produce reasonable answers. 

\begin{figure}
    \centering
    \includegraphics[width=\columnwidth]{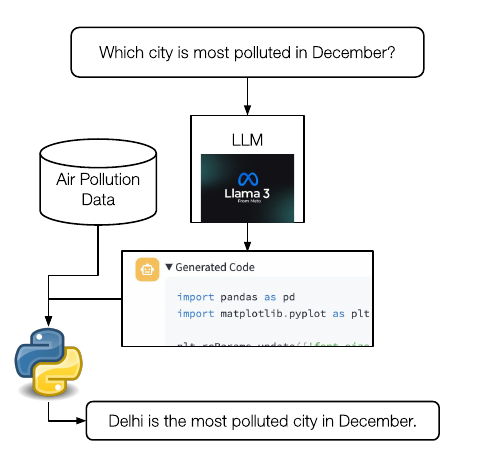}
    \caption{Flow Diagram of our \system\ System.}
    \label{fig:main}
\end{figure}

In this work, we present our system called  \system\footnote{https://huggingface.co/spaces/SustainabilityLabIITGN/VayuBuddy} which aims to provide an interface to ask questions about air pollution in natural language and get the answer in natural language or visual format. \system\ uses LLMs to convert natural language user queries into a Python code. The code is then executed by Python interpreter which has access to air quality sensor data. The final answer produced by the code is also in natural language. We use the Central Pollution Control Board (CPCB) air quality sensor data, specifically focusing on daily average air quality data, such as pollutant concentrations. Figure~\ref{fig:main} shows the flow diagram of our system.

We employed \nLLMs\ different LLMs on \nPrompts\ curated prompts, focusing on stakeholder needs. We generate these questions based on discussions with an air quality expert. 
To evaluate the system performance, we curate queries that can cover a variety based on: i) output type (plot/text); ii) query hardness; iii) query type (spatial, temporal, etc.). 
We add a common system prompt across LLMs to contextualize questions. For example, we mentioned the permissible pollution limits in India and requested the system to show them in the plot while plotting any time-series data. We can see those thresholds as dotted lines in Figure~\ref{App Screenshot}(d).

We ensured code reproducibility to facilitate transparency and enable others to replicate our methodology and results. Our web application can be found at here\footnote{\url{https://ouranonymoussubmission-vayubuddy.hf.space/}}. We believe that our work will lower the barrier into automatic air quality sensor data analytics for various stakeholders and overall help reduce the air pollution for social good.

\section{Related Work}

\subsection{Chatbots}
Chatbots have gained significant attention in a last few years. Various approaches have been explored to develop chatbots for different domains, including healthcare, customer service, education, and environmental monitoring~\cite{christoforakos2021connect}. In addition to offering real-time communication, chatbots have the potential to inform people about safety precautions, health dangers, and environmental data~\cite{10417387}. Our chatbot system aims to serve various stakeholders such as new media reporters, parents, lung patients, think tanks, policymakers, air quality researchers and general public.

\subsection{Codegen LLMs}
`Codegen LLM' is an LLM which can generate code for various programming languages. In numerous programming tasks, specialized Codegen LLMs—such as OpenAI's Codex\footnote{\href{https://openai.com/index/openai-codex}{https://openai.com/index/openai-codex}}, perform well and produce readable and accurate Python code~\cite{10.1145/3545945.3569830}. On the other hand, open-source models like CodeLlama~\cite{roziere2023code} has been performing at par with the closed source LLMs. Recent general-purpose LLMs are also showing promising code-generation capabilities. We benchmark both general-purpose and code-gen LLMs against hand-curated query-answer dataset created by us.

\subsection{LLMs for tabular data}
Language models (LLMs) tailored for tabular data processing have emerged as powerful tools for handling structured data in a natural language format. These models are designed to understand and generate text representations of tabular data, enabling tasks such as data summarisation, querying, and analysis. Prior research has demonstrated the effectiveness of LLMs for tabular data in domains such as finance, healthcare, and e-commerce~\cite{10.1145/3616855.3635752, yang2024unleashing}. We emphasize that LLMs which work directly on data are likely to have performance bottlenecks dependent on the size of the data. On the other hand, code-generation LLMs do not need to access the data itself so their performance is independent of the dataset size.

\subsection{Air Quality Toolkits}
Libraries, such as Vayu\footnote{\href{https://sustainability-lab.github.io/vayu/}{https://sustainability-lab.github.io/vayu/}} (for Python) and OpenAir\footnote{\href{https://www.openair.com}{https://www.openair.com}} (for R), help users visualize air-quality data and to generate meaningful insights. Our chatbot acts like a dynamic visualization library which does not require to be developed explicitly with software engineering. To improve the performance, it's possible to fine-tune the chatbot on hand-generated query-code pairs.

\section{Dataset}
\subsection{Air Quality Dataset}
Our air quality dataset comprises pollution measurements of PM$_{2.5}$ concentrations. The data is curated from 537 continuous monitoring stations from 279 cities across 31 states installed by the Central Pollution Control Board (CPCB), India. PM$_{2.5}$ is the most hazardous pollutant to health and thus we primarily focuses on PM2.5 in this work. We have included 7 years of data from 2017 to 2023. The data is originally collected every 15 minutes, however for computational efficiency, we resample it to daily.




\subsection{Natural Language Prompts Dataset}

We evaluate the performance of our system using two datasets: one focused on system capabilities and the other on stakeholder needs. The strategies for each dataset are as follows:

\begin{itemize}
\item System End: This dataset includes prompts created by authors and air quality experts to evaluate the capability of the system. We categorized the prompts based on the type of output they generate, such as numerical answers, textual answers or plots. Additionally, we differentiated prompts by the nature of the queries. 
Sample questions of such categories are presented in Table~\ref{Example Prompts}.

\item Stakeholder End: To create this dataset, we include the questions addressing the needs of various stakeholders. With input from an air quality expert, we collected several questions directly from them. For the remaining questions, we thought from the perspectives of various stakeholders. For example, policymakers may seek insights into overall pollution trends and compliance with the air quality standards, while patients might inquire about relocating to a city with lower pollution levels. Parents may want to understand the air pollution exposure of their children during weekdays. Sample questions from various stakeholders are presented in Table~\ref{tab:queries}.
\end{itemize}

\begin{table}[ht]
\centering
\centering
\begin{tabular}{p{0.3\columnwidth}p{0.62\columnwidth}}
\toprule
\textbf{Category (Count)} & \textbf{Example Prompt}\\
\midrule
Plot Output (12) & Plot the yearly average PM2.5.\\
\hline
Text Output (8) & Which city has the highest PM2.5 level in July 2022?\\
\hline
Spatial (11) & Which state has the highest average PM2.5?\\
\hline
Temporal (20) & Plot the monthly average PM2.5 of Delhi.\\
\hline
Raw Time (3) & Which city witnessed the lowest PM2.5?\\
\hline
Aggregated Time (17) & Plot the monthly average PM2.5 of Delhi.\\
\hline
Easy (2) & Plot the yearly average PM2.5.\\
\hline
Moderate (12) & Which month in which year has the highest PM2.5 overall?\\
\hline
Complex (6) & Plot and compare the monthly average PM2.5 of Delhi, Mumbai and Bengaluru for the year 2022.\\
\bottomrule
\end{tabular}
\vspace{10pt}
\caption {Examples of the prompts that can generate a wide variety of outputs at various difficulty levels. We include a mix of textual, numerical and visual outputs.}
\label{Example Prompts}
\end{table}

Following this procedure, we compiled a diverse set of 45 questions for evaluation. For each question, we manually wrote the corresponding Python code to retrieve accurate answers from the dataset. 

\begin{table}[ht]
\centering
\centering
\begin{tabular}{p{0.3\columnwidth}p{0.62\columnwidth}}
\toprule
\textbf{Category} & \textbf{Example Prompt}\\
\midrule
Policymakers & Number of cities that had PM2.5 levels above the WHO guideline in November 2023?\\
\hline
Air Quality Researcher & Which season (Summer, Winter, Spring, Autumn) experiences highest pollution levels?\\
\hline
Lung Patients & Which city in India has the best air quality?\\
\hline
Parents & What is the average air pollution on the weekdays?\\
\hline
Public & Which city has the highest PM2.5 level in July 2022?\\
\bottomrule
\end{tabular}
\vspace{10pt}
\caption {Example queries involving perspectives and needs of various stakeholders.}
\label{tab:queries}
\end{table}




\subsection{Large Language Models}
We compare \nLLMs\ Open-source LLM models in this work. We didn't use any paid LLM to keep the experiments reproducible to the research communities may or may not having the budget to spend on subscribing the LLMs. We describe each of the LLMs in brief in the following subsections.

\begin{itemize}
\item \textbf{Llama}: Llama-series LLMs from Meta have model sizes of 8 and 72 billion parameters with a context window of 8192 tokens. They included a tokenizer with a vocabulary of 128K tokens for more efficient language encoding and the adoption of grouped query attention (GQA) to enhance inference efficiency. The models are trained on sequences of 8192 tokens with masking to prevent self-attention from crossing document boundaries. \textbf{We have also included a very recently launched model LLama3.1 from Meta AI}.

\item \textbf{Mixtral}: This is a sparse mixture-of-experts ~\cite{sanseviero2023moe} network with a model size of 7B (known as Mistral) and 56 billion parameters and a context window of 32768 tokens. It utilizes only 13 billion active parameters during inference since it is a sparse mixture of networks. The model consists of experts, which are feed-forward neural networks, and a gate network or router that determines the routing of tokens to different experts. This router is pre-trained alongside the rest of the network.

\item \textbf{Codestral}: Codestral is code specific model from Mistral AI group with 22B parameters. It is trained on 80+ programming languages including C, Java and Python.

\item \textbf{Gemma}: A model with the size of 7 billion parameters and a context window of 8192 tokens adopts key improvements such as Multi-Query Attention, RoPE Embeddings, GeGLU Activations, and RMSNorm. The training process begins with supervised fine-tuning on a mix of text-only, English-only synthetic, and human-generated prompt-response pairs, followed by reinforcement learning from human feedback (RLHF). The reward model is trained on labelled English-only preference data, and the policy is based on a set of high-quality prompts.

\end{itemize}
 
\subsection{Prompt Engineering}\label{Customisation}
Note that LLM models needs to be given some well-engineered system prompts to get the most out of them. It can include but is not limited to the type of task, metadata about the questions, static information (for example, WHO and India limits of PM$_{2.5}$.) and answer formats. We provide the metadata of the dataset, static information and answer formats in system prompts. We describe the provided system prompt in Listing~\ref{sysprompt2}.

\lstset{
  basicstyle=\ttfamily,
  columns=fullflexible,
  frame=single,
  backgroundcolor=\color{lightgray}, 
  rulecolor=\color{darkgray}, 
  breaklines=true,
  breakindent=0pt, 
  tabsize=2, 
}


\begin{lstlisting}[mathescape=true, caption=System Prompts. Additional information and direction given to the LLM., label=sysprompt2]
- The columns are 'Timestamp', 'station', 'PM2.5', 'PM10', 'address', 'city', 'latitude', 'longitude',and 'state'.
- Frequency of data is daily.
- `pollution` generally means `PM2.5`.
- Don't print anything, but save result in a variable `answer` and make it global.
- Unless explicitly mentioned, don't consider the result as a plot.
- PM2.5 guidelines: India: 60, WHO: 15.
- PM10 guidelines: India: 100, WHO: 50.
- If result is a plot, show the India and WHO guidelines in the plot.
- If result is a plot make it in tight layout, save it and save path in `answer`. Example: `answer='plot.png'`
- If result is a plot, rotate x-axis tick labels by 45 degrees.
- If result is not a plot, save it as a string in `answer`. Example: `answer='The city is Mumbai'`
- I have a geopandas.geodataframe india containining the coordinates required to plot Indian Map with states.
- If the query asks you to plot on India Map, use that geodataframe to plot and then add more points as per the requirements using the similar code as follows : v = ax.scatter(df['longitude'], df['latitude']). If the colorbar is required, use the following code : plt.colorbar(v)
- If the query asks you to plot on India Map plot the India Map in Beige color 
- Whenever you do any sort of aggregation, report the corresponding standard deviation, standard error and the number of data points for that aggregation.
- Whenever you're reporting a floating point number, round it to 2 decimal places.
- Always report the unit of the data. Example: The average PM2.5 is 45.67 $\mu gm^{-3}$.
\end{lstlisting}

\section{Evaluation}\label{Eval}
We now evaluate the efficacy of the different LLMs on the different prompts.


\subsection{Experimental Setup}
We created an automated evaluation pipeline to accelerate the evaluation process. We ensured fairness in the evaluation process by incorporating all the possible ways of correct answers in the pipeline. We measured the correctness in the following manner: A score of 1 is assigned if the response obtained is correct. Any other response is treated as incorrect and scored 0. There are three possible ways for a response to be considered incorrect:
\begin{enumerate}
    \item LLM failing to generate the code.
    \item The LLM generates the code, but the code gives error while executing.
    \item The LLM generated code runs without errors but gives a wrong answer.
\end{enumerate}

We use all the default configurations for the LLMs except temperature, which is set to 0 to disable randomness in the answers.

\subsection{Results}

In this section we present the results from evaluation on various LLMs. Table~\ref{tab:overall} shows the overall scores of LLMs across all categories. We observe that LLama3.1, which is a recently released open-source model by Meta, is gaining the best score. However it is just 1 point ahead of its predecessor, LLama3. Codestral and Mixtral models are also reasonably good but perhaps needs a more specific prompt engineering to increase the performance.

\begin{table}[ht]
\centering

\begin{tabular}{lcr}
\toprule
LLM &   \# params & Score (out of \nPrompts) \\
\midrule
Llama3.1        & 70B &  39 \\
Llama3          & 70B & 38 \\
Codestral       & 22B & 29 \\
Mixtral         & 56B & 26 \\
Llama3.1          & 8B & 23\\
Llama3          & 8B & 21\\
Gemma & 9B & 19\\
Codestral Mamba & 7B & 19 \\
Mistral         & 7B & 8 \\
Gemma           & 7B & 7 \\

\bottomrule
\end{tabular}
\vspace{10pt}
\caption {Overall Performance of LLMs on all evaluation queries. LLama3 models are performing the best among all models. Codestral and Mixtral are following LLama in performance. Models with more parameters are performing well compared to less parameters.}
\label{tab:overall}
\end{table}



\begin{table}[ht]
\renewcommand{\arraystretch}{1.2}
\centering
\begin{tabular}{lrrrrr}
\toprule
{LLM} &  1 &  2 &  3 &  4 &  5 \\
\midrule
Llama3-70b        & 19 & 15 & 16 & 20 & 24 \\
Mixtral           & 15 & 11 & 11 & 15 & 14 \\
Gemma-7b          &  2 &  2 &  6 &  6 &  4 \\
Llama3.1-70b      & 20 & 16 & 16 & 20 & 24 \\
Codestral Mamba   &  8 &  7 &  8 &  9 & 13 \\
Codestral         & 14 & 12 & 15 & 18 & 19 \\
Mistral 7B        &  5 &  3 &  3 &  3 &  5 \\
Llama3-8b         & 11 &  7 & 11 & 14 & 15 \\
Llama3.1-8b       & 10 &  8 & 12 & 13 & 17 \\
Gemma-9b          &  8 &  7 & 12 & 13 & 12 \\
\bottomrule
\end{tabular}
\vspace{10pt}
\caption {Category Wise performance of LLMs. 1 - Policymakers, 2 - AQ Researcher, 3 - Lung Patients, 4 - Parents, 5 - Public}
\label{Category wise Comparison Results}
\end{table}

\subsubsection{Analysis}
\begin{figure}
  \includegraphics[width=1\columnwidth]{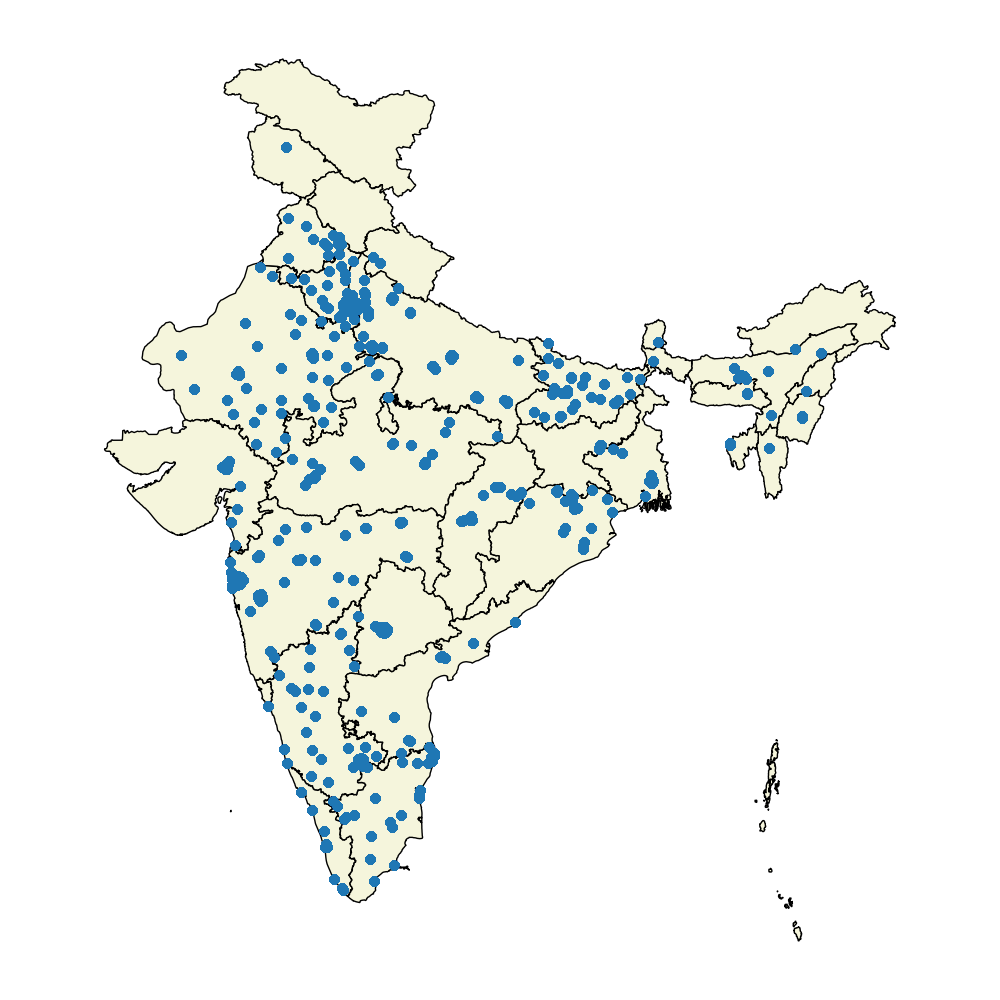}
  \caption{Location of Sensors: This image was generated by \system\ with the following prompt: "Plot the locations of the stations on the India Map. Do not Annotate."}
  \label{stations}
\end{figure}

\begin{figure}
  \includegraphics[width=1\columnwidth]{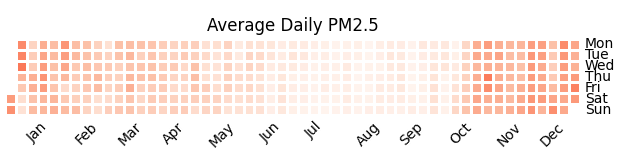}
  \caption{This image was generated by \system\ with the following prompt: "Create a calendar map showing average PM2.5."}
  \label{calander map}
\end{figure}

\begin{figure}
  \includegraphics[width=1\columnwidth]{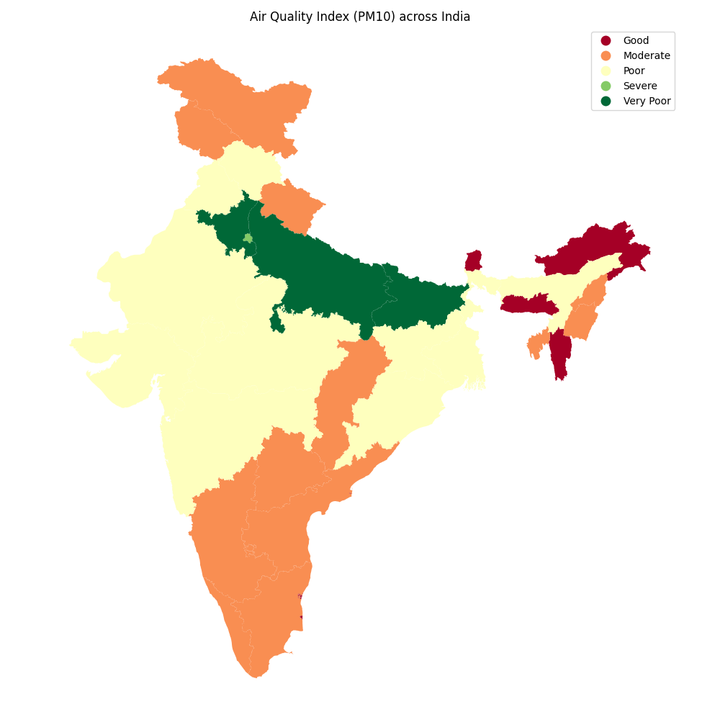}
  \caption{This image was generated by \system\ with the following prompt: "Plot the choropleth map showing PM10 levels across India, with different colours representing different AQI categories (e.g., Good, Moderate, Unhealthy, etc.)"}
  \label{chloropleth map}
\end{figure}

We observe the following points while analysing the results from various LLMs.
\begin{itemize}
   \item We observe that in almost all cases LLMs were able to generate either errorless or faulty Python codes. We rarely see a case where any code is not generated.
    \item Llama3 provides a good balance between code generation and general knowledge. Code based LLMs failed at questions which required prior information about lockdown and festival seasons, while models Gemma and Mistral lack pretraining on codes.
    \item Few questions like "How many days in 2023 did Mumbai exceed the WHO's PM2.5 guidelines?" could not be handled by any of the LLM due to the lack of proper extraction of feature based knowledge that each city can have multiple stations. 
    \item Gemma and Mistral mostly generate non-working Python codes which run into errors. This could be attributed to their small parameter size compared to other LLMs.
    \item We showcase three non-trivial prompts (Figure~\ref{stations}, Figure~\ref{calander map}, and Figure~\ref{chloropleth map}) and their corresponding plots generated correctly by \system. These show the richness of the capabilities.
\end{itemize}

\section{Limitations and Future Work}

\begin{itemize}
    \item \textbf{Expanding across countries: } In the current work, we looked at the air quality from India. In the future, we plan to expand by including air quality data from other countries. We plan to source the data from OpenAQ\footnote{\url{https://openaq.org/}} platform. Unfortunately due to some changes in the CPCB operation, OpenAQ does not presently provide Indian pollution data. Going beyond a single country's data might introduce additional challenges, and we may need to introduce a meta model to route to a model for a specific country. 
    \item \textbf{Expanding to text data: }
    We aim to enhance \system's capabilities to additionally answer queries based on text inputs. These could include various advisories from pollution control boards. 
    \item \textbf{Expanding to additional output types:}  In our initial testing we found that we can also generate plots from our \system\ system. However, evaluating the correctness of such plots would need a purpose-built rubric. We plan to study such output types in the future. 
    \item \textbf{Extending Feature Categories: }We plan to include more categories of features such as other pollutants like SO$_2$, NO$_2$, NO, CO, and Ozone. Along with that, we plan to add important meteorological parameters such as wind (direction and speed).
    

    \item \textbf{Enhancing Prompting Methods: }Our work currently relies on zero shot prompting, which involves prompting the model without specific examples. In the future, we plan to evaluate other strategies like chain of thought, tree of thought, and react prompting. These approaches encourage deeper consideration before generating responses, potentially refining \system's ability to offer nuanced insights on air quality-related queries. 
    \item \textbf{Finetuning:} Our current work has looked into zero-shot performance. In the future, we plan to study the finetuning performance of our models. 
    \item \textbf{Implementing Active Learning Strategies: }Another innovative approach for enhancing \system's capabilities is exploring active learning strategies. By integrating active learning methodologies, \system\ can intelligently select a set of prompts, optimising the training process and improving model performance over time. Active learning techniques can enhance \system's adaptability to new data and user interactions.
    \item \textbf{Automating Library Installation: }To streamline the user experience and enhance the autonomy of \system, a methodology for automated library installation could be developed. Instead of requiring users/us to manually install libraries and provide them to the model, \system\ could autonomously identify the necessary libraries based on the user's query and install them as needed. This solution would simplify the user interaction process and improve the efficiency and accessibility of \system\, allowing users to seamlessly access air quality insights without the burden of manual setup.

\end{itemize}

\section{Conclusion}
In this work, we explored our system \system to respond to natural language queries pertaining to air pollution based on data from CPCB. We believe that a system such as ours can be used by various stakeholders including but not limited to pollution control boards, researchers. Such systems put the focus back on the problem or the question being asked rather than the engineering efforts towards solving the query to extract the data. In our initial discussions with various air quality experts, the response has been positive. We plan to roll this to more stakeholders.


\bibliographystyle{ACM-Reference-Format}
\bibliography{main}
\end{document}